\documentclass[referee,twocolumn, times, linenumbers, tighten, a4paper,fleqn,usenatbib,useAMS]{raa_twocolumn}           
\usepackage{graphicx,times}
\usepackage{natbib}
\usepackage{xcolor}
\usepackage{mathtools}
\usepackage{adjustbox}
\usepackage[version=4]{mhchem}
\usepackage{amssymb,amsmath}
\bibpunct{(}{)}{;}{a}{}{,}

\usepackage[colorlinks=true, allcolors=blue]{hyperref}

\begin{document}

   \title{Study of Complex Nitrogen and Oxygen-bearing Molecules toward the High-mass Protostar IRAS 18089--1732}

 \volnopage{ {\bf 20XX} Vol.\ {\bf X} No. {\bf XX}, 000--000}
   \setcounter{page}{1}

   \author{Arijit Manna$^{*}$
   \inst{1}, Sabyasachi Pal\inst{1}, Tapas Baug\inst{2}, Sougata Mondal\inst{1}}

   \institute{ $^{1}$Department of Physics and Astronomy, Midnapore City College, Paschim Medinipur, West Bengal, India 721129, {\it Email: arijitmanna@mcconline.org.in}\\
   $^{2}$S. N. Bose National Centre for Basic Sciences, Block-JD, Sector-III, Salt Lake City, Kolkata, India 700106\\
        %\and
          %   Yunnan Astronomical Observatory, National Astronomical Observatories, Chinese Academy of Sciences,
          %   Kunmin 650011, China\\
	%\and
%	  Center for Astrophysics, University of Science and Technology of China, Hefei 230026, China\\
%Key Laboratory for Research in Galaxies and Cosmology, The University of Science
%and Technology of China, Chinese Academy of Sciences, Hefei, Anhui, 230026, China\\
%\and 
%Polar Research Institute of China,
%Jinqiao Rd. 451, Shanghai, 200136, China\\
\vs \no
   {\small Received 20XX Month Day; accepted 20XX Month Day}
}

\abstract{The observation of oxygen (O)- and nitrogen (N)-bearing molecules gives an idea about the complex prebiotic chemistry in the interstellar medium (ISM). Recent millimeter and submillimeter wavelength observations have shown the presence of complex O- and N-bearing molecules in the star-formation regions. So, the investigation of those molecules is crucial to understanding the chemical complexity in the star-forming regions. In this article, we present the identification of the rotational emission lines of N-bearing molecules ethyl cyanide (\ce{C2H5CN}), cyanoacetylene (\ce{HC3N}), and O-bearing molecules methyl formate (\ce{CH3OCHO}) towards high-mass protostar IRAS 18089--1732 using the Atacama Compact Array (ACA). We also detected the emission lines of both N- and O-bearing molecule formamide (\ce{NH2CHO}) in the envelope of IRAS 18089--1732. We have detected the $v$ = 0 and 1 states rotational emission lines of \ce{CH3OCHO}. We also detected the two vibrationally excited states of \ce{HC3N} ($v$7 = 1 and $v$7 = 2). The estimated fractional abundances of \ce{C2H5CN}, \ce{HC3N} ($v$7 = 1), \ce{HC3N} ($v$7 = 2), and \ce{NH2CHO} towards the IRAS 18089--1732 are (1.40$\pm$0.5)$\times$10$^{-10}$, (7.5$\pm$0.7)$\times$10$^{-11}$, (3.1$\pm$0.4)$\times$10$^{-11}$, and (6.25$\pm$0.82)$\times$10$^{-11}$. Similarly, the estimated fractional abundances of \ce{CH3OCHO} ($v$ = 0) and \ce{CH3OCHO} ($v$ = 1) are (1.90$\pm$0.9)$\times$10$^{-9}$ and (8.90$\pm$0.8)$\times$10$^{-10}$, respectively. We also created the integrated emission maps of the detected molecules, and the observed molecules may have originated from the extended envelope of the protostar. We show that \ce{C2H5CN} and \ce{HC3N} are most probably formed via the subsequential hydrogenation of the \ce{CH2CHCN} and the reaction between \ce{C2H2} and CN on the grain surface of IRAS 18089--1732. We found that \ce{NH2CHO} is probably produced due to the reaction between \ce{NH2} and \ce{H2CO} in the gas phase. Similarly, \ce{CH3OCHO} is possibly created via the reaction between radical \ce{CH3O} and radical HCO on the grain surface of IRAS 18089--1732.
}

   \authorrunning{Manna et al.}            %author_head in even pages
   \titlerunning{Complex organic molecules towards IRAS 18089--1732}  % title_head in odd pages
   \maketitle

%________________________________________________ sections below
% 
\section{Introduction}
The study of complex organic molecules (COMs) from massive stars ($M$ $>$ 8 $M_{\odot}$) is important for understanding the chemical complexity in the interstellar medium (ISM) \citep{van98, her09}. High-mass stars generate a significant amount of heavy atomic elements, produce large amounts of UV radiation, and implement turbulent energy into the ISM \citep{zap06}. High-mass protostars are located in a rich cluster environment, and it is challenging to understand the physical and chemical formation processes \citep{de05}. High-mass protostars are highly luminous ($\ge$10$^{3}$ L$_{\odot}$), and these sources are deeply embedded in massive envelopes that contain a high gas temperature ($\ge$100 K) and gas density above 10$^{6}$ cm$^{-3}$ \citep{li21}. The study of the physical phenomena of high-mass protostellar disks is a major concern in high-mass star-forming regions. Over the past few years, indirect evidence for the existence of massive disks has accumulated, but there is still no clear proof of massive disks. Additionally, there are several methods for directly studying massive disks. Infrared and millimeter-wavelength continuum observations discover several attractive disk candidates, but they are unable to provide kinematic proof of potential accretion disks owing to a lack of velocity information \citep{sh00, ch04}. On the other hand, class II methanol (\ce{CH3OH}) and water (\ce{H2O}) maser emission found kinematic signatures from the massive disk candidates, but since maser emission is extremely selective and demands specific conditions, such studies do not permit more statistical analysis of massive disk properties \citep{tor96, pes04}. The most likely method to find and investigate disks in high-mass star formations is to detect thermal molecular line emissions, which are predicted to be sensitive to the gas properties and kinematics of the disks. Therefore, spectral line surveys in the millimeter and sub-millimeter wavelengths are crucial for understanding molecular gas properties, such as column density and excitation temperature. The best method is to search the rotational emission lines of acetonitrile (\ce{CH3CN}) and methyl acetylene (\ce{CH3CCH}) because both molecules are known as gas thermometers for interstellar gas in massive protostars, hot molecular cores, and hot corinos \citep{and18}. The prebiotic chemistry of high- and low-mass protostellar candidates may be the same, and a detailed analysis of the chemical evolution of high- and low-mass protostellar sources is not only important for astrochemistry but also crucial for understanding the process of star formation. Over the past decade, a variety of COMs have been discovered in protostellar cores on the scale of a few thousand astronomical units \citep{sak13}. Two different scenarios were identified: (1) prebiotic chemistry in hot corinos, which is characterized by the richness of saturated COMs \citep{caz03, her09, cas12}, and (2) warm carbon chain chemistry, which is characterized by the richness of unsaturated hydrocarbons \citep{sak08}. Studying the chemical evolution of protostellar sources requires the study of isolated sources that are unaffected by nearby protostellar feedback. Isolated protostellar sources are excellent laboratories for theories of both chemical evolution and star formation \citep{ev15}.

\begin{figure*}
\centering
\includegraphics[width=1.0\textwidth]{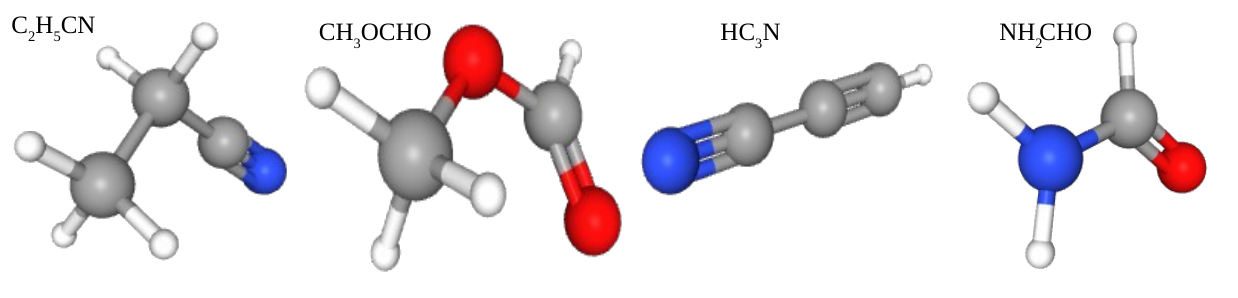}
\caption{Three-dimensional molecular structure of \ce{C2H5CN}, \ce{CH3OCHO}, \ce{HC3N}, and \ce{NH2CHO}. The grey atoms are carbon (C), the red atoms are oxygen (O), the white atoms are hydrogen (H), and the blue atoms are nitrogen (N).}
\label{fig:proto}
\end{figure*}

\begin{figure}
\centering
\includegraphics[width=0.8\textwidth]{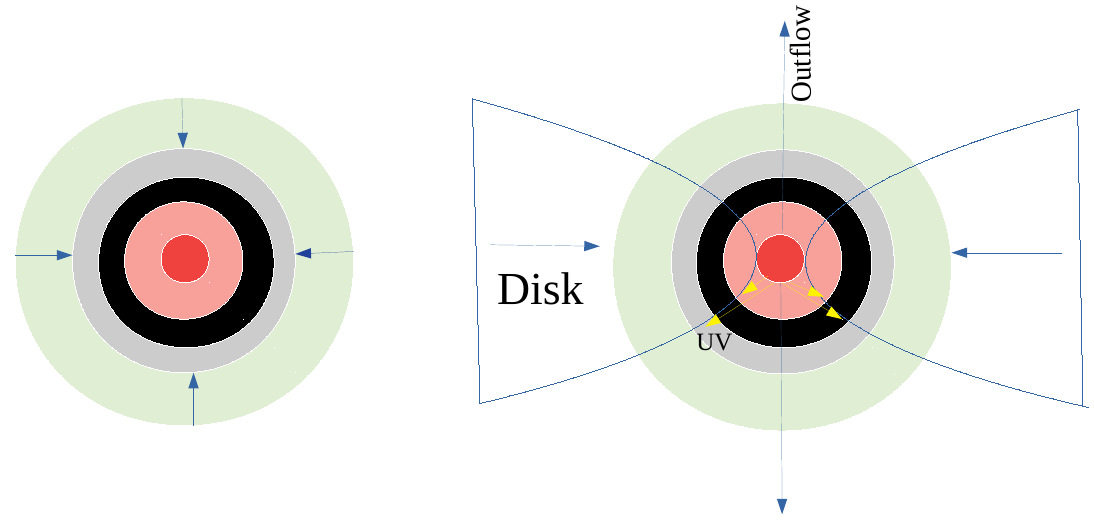}
\caption{A cartoon diagram of a high-mass protostar with a spherical shape (left) and a protostar with a flattened disk-like structure (right), with increasing UV photons illuminating the walls of the outflow cavity.}
\label{fig:dia}
\end{figure}

Most chemical models of protostellar sources and hot molecular cores utilize grain surface chemistry to produce different types of complex organic compounds \citep{ti97, gar13, man24}. The hydrogenation of solid oxygen (O), carbon (C), nitrogen (N), and carbon monoxide (CO) in the cold ($\leq$20 K) prestellar phase produces \ce{CH3OH} and other hydrogenated species such as \ce{H2O}, \ce{NH3}, and \ce{CH4} \citep{ti82}. Ultraviolet radiation photodissociates these surface ices, and when the cloud core heats up during the protostellar phase, the fragments become mobile. Consequently, complex molecules are formed during the subsequent recombination of the photofragments and evaporate when the grain temperature increases above the ice sublimation temperature of $\geq$100 K \citep{gar06, gar08, man24}. For example, \ce{CH3OCH3}, \ce{CH3OCHO}, and \ce{C2H5OH} are formed by UV processing of \ce{CH3OH} ice \citep{ob09}. Finally, the hot-core gas phase chemistry between evaporated molecules may cause second-generation species to become more complex \citep{mil91, cha92}. One of the most significant impacts of an equatorial rather than a spherical structure is that UV light may more easily exit from the central source and illuminate the surface layers of the surrounding disk or toroid, as well as the larger-scale envelope \citep{bru09, bru10, iso13} (Figure~\ref{fig:dia}). This can result in ice-producing COMs that are faster than \ce{CH3OH}. Another effect of UV radiation is enhanced photodissociation of gaseous \ce{N2} and CO. Consequently, more atomic N and C are easily accessible for grain surface chemistry, which might result in an increased abundance of species such as HNCO and \ce{NH2CHO} \citep{gar13}.

IRAS 18089--1732 (hereafter IRAS 18089) is a well-known high-mass protostar that contains an ultra-compact (UC) H{\sc ii} region \citep{li21}. That source is located at a distance of 2.34 kpc from the Earth \citep{xu11}. The luminosity and gaseous mass of IRAS 18089 are 1.3$\times$10$^{4}$ $L_{\odot}$ and 1000 $M_{\odot}$, which is estimated from the single-dish millimeter wavelength continuum emission \citep{sri02, beu02a}. Earlier, the emission lines of \ce{CH3OH}, SO, {SO$_{2}$}, {H$_{2}$S}, and many other COMs were detected towards the IRAS 18089, but no column densities or abundances were reported \citep{beu04, iso13}. Earlier estimated excitation temperature of \ce{CH3CN} ($\sim$ 350 K) towards IRAS 18089 indicated that the source deeply embedded the hot core \citep{beu04}. The molecular outflow of SiO (J = 5--4) is also evident towards the IRAS 18089, which was detected using the Submillimeter Array (SMA) \citep{beu04}. Additionally, the Goldreich-Kylafis effect was observed for the CO (J = 3--2) emission line, which revealed a linear polarization fraction of up to 8\% \citep{beu10}. The {H$_{2}$O} and {CH$_{3}$OH} maser emissions were also detected towards the IRAS 18089 \citep{beu02b}. Recently, \citet{qi22} made a spectral line survey of the rotational emission lines of ethyl cyanide (\ce{C2H5CN}), methyl formate (\ce{CH3OCHO}), and methanol (\ce{CH3OH}) towards the 146 high-mass star-forming regions using the ATOMS survey data \citep{li20}. \citet{qi22} mentioned the detection of only two and six transition lines of \ce{C2H5CN} and \ce{CH3OCHO} towards IRAS 18089 among which one transition line of \ce{C2H5CN} and two transition lines of \ce{CH3OCHO} are blended with other molecules (see supplementary data of \citet{qi22}). \citet{qi22} also does not discuss the abundance and proper gas phase and grain surface chemistry of those molecules towards IRAS 18089. This indicates that another detailed molecular spectral line study is needed to understand the abundance and prebiotic chemistry of different COMs towards IRAS 18089.

In this paper, we present the detection of rotational emission lines of \ce{C2H5CN}, \ce{CH3OCHO}, \ce{HC3N}, and \ce{NH2CHO} towards the IRAS 18089. The chemical diagram of those molecules is created using the molecular modelling software {\tt Avogadro} \citep{han12} is shown in Figure~\ref{fig:proto}. Previously, the emission lines of \ce{C2H5CN}, \ce{CH3OCHO}, \ce{HC3N}, and \ce{NH2CHO} was detected towards hot molecular cores G10.47+0.03 \citep{gor20, mon21, man23, mon23}, G31.41+0.31 \citep{min23}, Orion KL \citep{beu05a}, Sgr B2 (N) \citep{bel13}, and low-mass protostar IRAS 16293-2422 \citep{jo16, cal18}. The paper is organized in the following manner. The observations and data reduction are presented in Section~\ref{sec:obs}. The result of the detection of emission lines of complex molecules is shown in Section~\ref{res}. The discussion and conclusion of the detection of complex molecules are shown in Section~\ref{dis} and \ref{con}.

\begin{table*}{}
\centering
\caption{Observation summary of IRAS 18089.}
\begin{adjustbox}{width=1.0\textwidth}
\begin{tabular}{cccccccccccc}
\hline
			Observation date &	Integration time&Frequency range&Spectral resolution &Sensitivity (10 km s$^{-1}$)&Field of view (FOV)\\	
			(yyyy-mm-dd)     &	(s)       &(GHz)            & (kHz)                 &(mJy beam$^{-1}$)&($^{\prime\prime}$)\\
			\hline		
			2017-09-18      &	5624.64       &127.49--128.48&488.28               &	5.91 &74.23 	\\
			--              &   --           &129.75--130.75 &488.28                            &5.37&--\\
			--              &   --           &139.09--140.09 &488.28   &5.15&--\\
			--               &   --           &140.45--141.45 &488.28     &5.01&--\\
			
			\hline
\end{tabular}	
\end{adjustbox}

	\label{tab:data}
	%\end{minipage}[t]{\columnwidth}
\end{table*}

\section{Observations and data reductions}
\label{sec:obs}
We used the cycle 4 archival data of IRAS 18089, observed using the 7-m Atacama Compact Array (ACA) of the Atacama Large Millimeter/Submillimeter Array (ALMA) (ID: 2016.2.00005.S. PI: Rivilla, Victor). The observations were performed in Band 4 (frequency range 127.49 GHz -- 141.45 GHz). The observed phase centre of IRAS 18089 was ($\alpha,\delta$)$_{\rm J2000}$ = 18:11:51.400, --17:31:28.00. The observation was carried out on September 18, 2017, using eight antennas. The observations were made with spectral ranges of 127.49--128.48 GHz, 129.75--130.75 GHz, 139.09--140.09 GHz, and 140.45--141.45 GHz, with a corresponding spectral resolution of 488 kHz. During the observation of IRAS 18089, the flux calibrator and bandpass calibrator were J1924--2914, and the phase calibrator was taken as J1833--210B. The observation summary is shown in Table~\ref{tab:data}.

We used the Common Astronomy Software Application (CASA 5.4.1) with the data reduction automated pipeline for data reduction and spectral imaging of IRAS 18089 \citep{mc07}. For flux calibration of IRAS 18089, we used the Perley-Butler 2017 flux calibrator model for each baseline to scale the continuum flux density of the flux calibrator using the CASA task {\tt SETJY} with 5\% accuracy \citep{per17}. Using the CASA pipeline tasks {\tt hifa bandpassflag} and {\tt hifa flagdata}, we create the flux and bandpass calibration after flagging the defective channels and antenna data. After the initial data reduction, we used the CASA task {\tt MSTRANSFORM} with all available rest frequencies to separate the target data of IRAS 18089. We created the continuum emission maps of IRAS 18089 for line-free channels using the CASA task {\tt TCLEAN} with {\tt HOGBOM} deconvolver. After the production of the continuum images, we used the task {\tt UVCONTSUB} in the UV plane of separated calibrated data of IRAS 18089 for the continuum subtraction. To create the spectral images of IRAS 18089, we used the CASA task {\tt TCLEAN} with {\tt SPECMODE = CUBE} parameter and Briggs weighting robust value of 0.5. Both the continuum map and spectral cube were corrected for the primary beam using the CASA {\tt IMPBCOR} task.

\section{Result}
\label{res}
\subsection{Dust continuum emission}
\begin{figure*}
	\centering
	\includegraphics[width=0.55\textwidth]{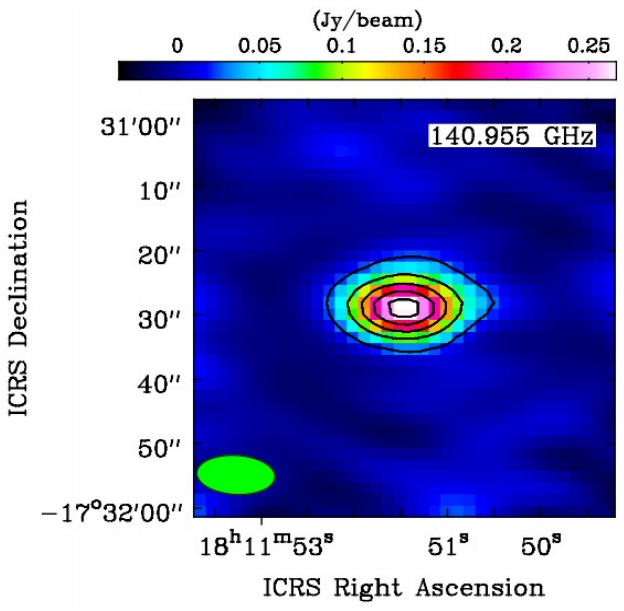}
	\caption{The 2.1 mm continuum map of the IRAS 18089 overlaid with black contours. The contour levels start at 3$\sigma$, where $\sigma$ is the RMS (26.63 mJy beam$^{-1}$) of the continuum image. The contour levels increase by a factor of $\surd$2. The green circle indicates the synthesised beam of the continuum image.}
	\label{fig:continuum}
\end{figure*}

The 2.1-mm dust continuum image of IRAS 18089 is shown in Figure~\ref{fig:continuum}. We employ the CASA {\tt IMFIT} task to estimate parameters of the central core, such as integrated flux density, peak flux density, corresponding position angle, and RMS noise level. The estimated peak flux density and integrated flux density of the source are 274.9$\pm$6.5 mJy beam$^{-1}$ and 346.1$\pm$14 mJy, respectively, with a RMS noise of 26.63 mJy beam$^{-1}$. The deconvolved source size is 5.14$''\times$3.59$''$ at a position angle of 89.8$^{\circ}$. The synthesised beam size of the continuum image is 12.78$^{\prime\prime}$$\times$7.20$^{\prime\prime}$. Thus, the continuum core is not resolved in the IRAS 18089 region.

\subsubsection{Estimation of hydrogen (H$_{2}$) column density}
For optically thin dust continuum emission, the peak flux density ($S_\nu$) can be expressed as\\

%\begin{equation}
$S_\nu = B_\nu(T_d)\tau_\nu\Omega_{beam}$ \citep{gor20} ~~~~~~~~~~~~~~~~~~~~~~~~~~~~~~~~~~~~~~~~~~~~~~~(1)\\\\
%\end{equation}
where $B_\nu(T_d)$ presents the Planck function at dust temperature $T_d$ \citep{whi92}, $\tau_\nu$ indicates the optical depth, and $\Omega_{beam} = (\pi/4 \ln 2)\times \theta_{major} \times \theta_{minor}$ indicates the solid angle of the synthesized beam. The equation of optical depth in terms of the mass density of dust can be written as,\\

$\tau_\nu =\rho_d\kappa_\nu L$ \citep{gor20}~~~~~~~~~~~~~~~~~~~~~~~~~~~~~~~~~~~~~~~~~~~~~~~~~~~~~~~~~~~~~~(2)\\\\
where $\rho_d$ presents the mass density of dust, $\kappa_{\nu}$ is the mass absorption coefficient, and $L$ defines the path length. The mass density of the dust can be expressed in terms of the dust-to-gas mass ratio ($Z$),\\

$\rho_d = Z\mu_H\rho_{H_2}=Z\mu_HN_{H_2}2m_H/L$ ~~~\citep{gor20}~~~~~~~~~~~~~~~~~~~~~~~(3)\\\\
where $\mu_H$ defines the mean atomic mass per hydrogen, $\rho_{H_2}$ is the hydrogen mass density, $m_H$ presents the mass of hydrogen, and $N_{H_2}$ is the column density of molecular hydrogen. We take the dust temperature $T_d$ = 30 K \citep{san21}, $\mu_H = 1.41$, and $Z = 0.01$ \citep{cox00}. The estimated peak flux density of the dust continuum of IRAS 18089 at frequency 140.955 GHz is 274.9$\pm$6.5 mJy beam$^{-1}$. From equations 1, 2, and 3, the column density of molecular hydrogen can be expressed as,\\

$N_{H_2} = \frac{S_\nu /\Omega}{2\kappa_\nu B_\nu(T_d)Z\mu_H m_H}$ ~~~\citep{gor20}~~~~~~~~~~~~~~~~~~~~~~~~~~~~~~~~~~~~~~~~~(4)\\\\
For estimation of the mass absorption coefficient ($\kappa_{\nu}$), we adopt the formula $\kappa_\nu = 0.90(\nu/230~\textrm{GHz})^{\beta}\ \textrm{cm}^{2}\ \textrm{g}^{-1}$ \citep{moto19}, where $k_{230} = 0.90$ cm$^{2}$ g$^{-1}$ indicates the emissivity of the dust grains at a gas density of $\rm{10^{6}\ cm^{-3}}$, which is covered by a thin ice mantle at 230 GHz. The dust spectral index ($\beta$) of $\sim$1.6 towards IRAS 18089 is adopted from \citet{san21}. Using the mass absorption coefficient formula, the estimated value of $\kappa_{\nu}$ is 0.411. Using equation 4, we estimate the column density of the molecular hydrogen ($N_{H_{2}}$) towards the IRAS 18089 to be (8.01$\pm$0.4)$\times$10$^{23}$ cm$^{-2}$. The earlier estimated column density of molecular \ce{H2} towards the IRAS 18089 at wavelength 1.2 mm using the SMA is 9.5$\times$10$^{23}$ cm$^{-2}$ \citep{beu05b}, which is nearly similar to our derived column density of \ce{H2} ((8.01$\pm$0.4)$\times$10$^{23}$ cm$^{-2}$) using the ACA at wavelength 2.1 mm.

\begin{figure*}
	\centering
	\includegraphics[width=0.97\textwidth]{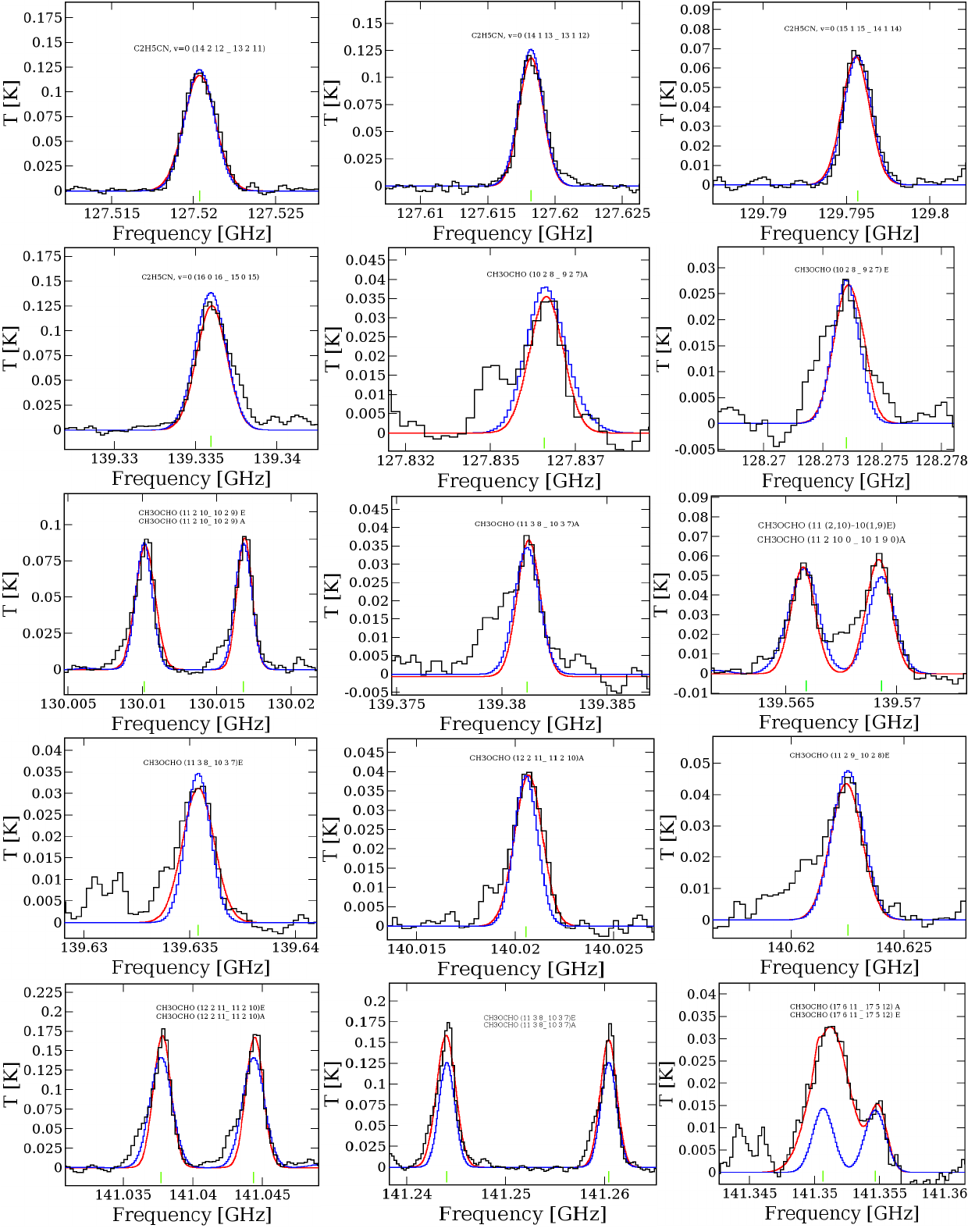}
	\caption{Rotational emission lines of \ce{C2H5CN}, \ce{CH3OCHO} (both V$_{t}$ = 0 and V$_{t}$ = 1), \ce{HC3N} (both $v$7 = 1 and $v$7 = 2), and \ce{NH2CHO} towards the IRAS 18089 with different transitions. The black lines present the observed millimeter-wavelength spectra of IRAS 18089 and the blue synthetic spectra present the LTE model spectra of \ce{C2H5CN}, \ce{CH3OCHO} (V$_{t}$ = 0 and V$_{t}$ = 1), \ce{HC3N} ($v$7 = 1 and $v$7 = 2), and \ce{NH2CHO}. The red spectrum indicates the Gaussian model which is fitted over the detected molecular spectra. The green vertical lines indicate the rest frequency positions of the detected transitions of \ce{C2H5CN}, \ce{CH3OCHO} (both V$_{t}$ = 0 and V$_{t}$ = 1), \ce{HC3N} (both $v$7 = 1 and $v$7 = 2), and \ce{NH2CHO}.}
	\label{fig:coms}
\end{figure*}
\begin{figure*}
	\text{{\large Figure~4 Continued.}}
	\centering
	\includegraphics[width=0.97\textwidth]{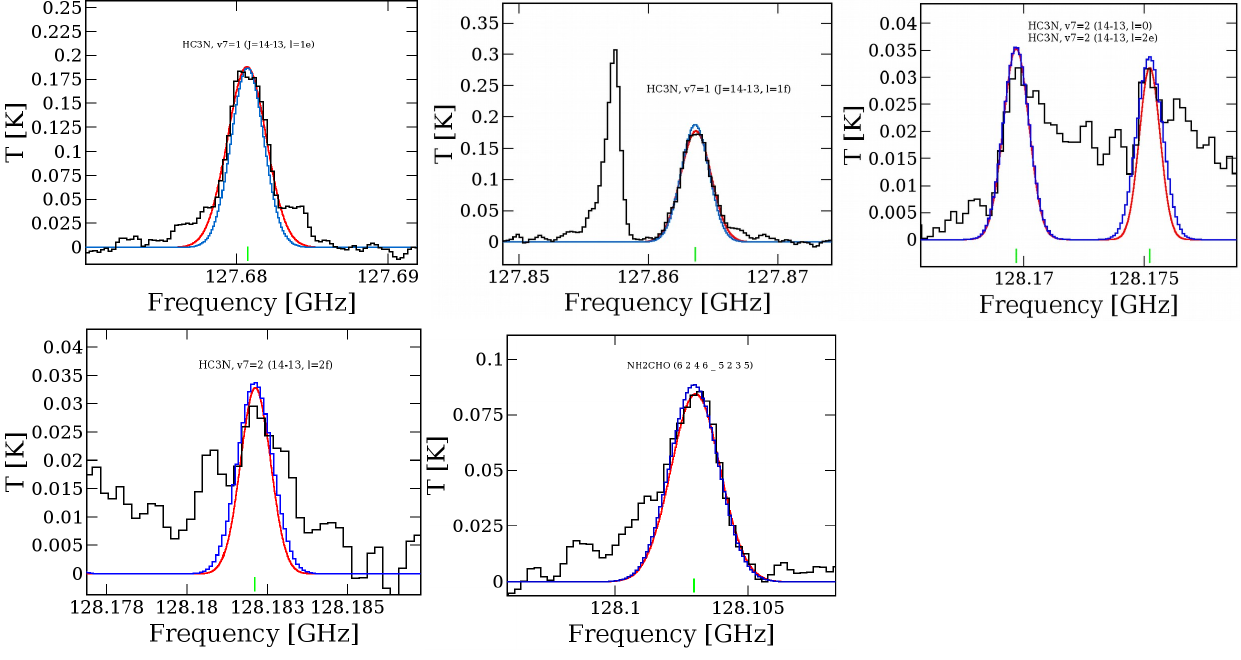}
\end{figure*}

\begin{figure*}
	\centering
	\includegraphics[width=0.95\textwidth]{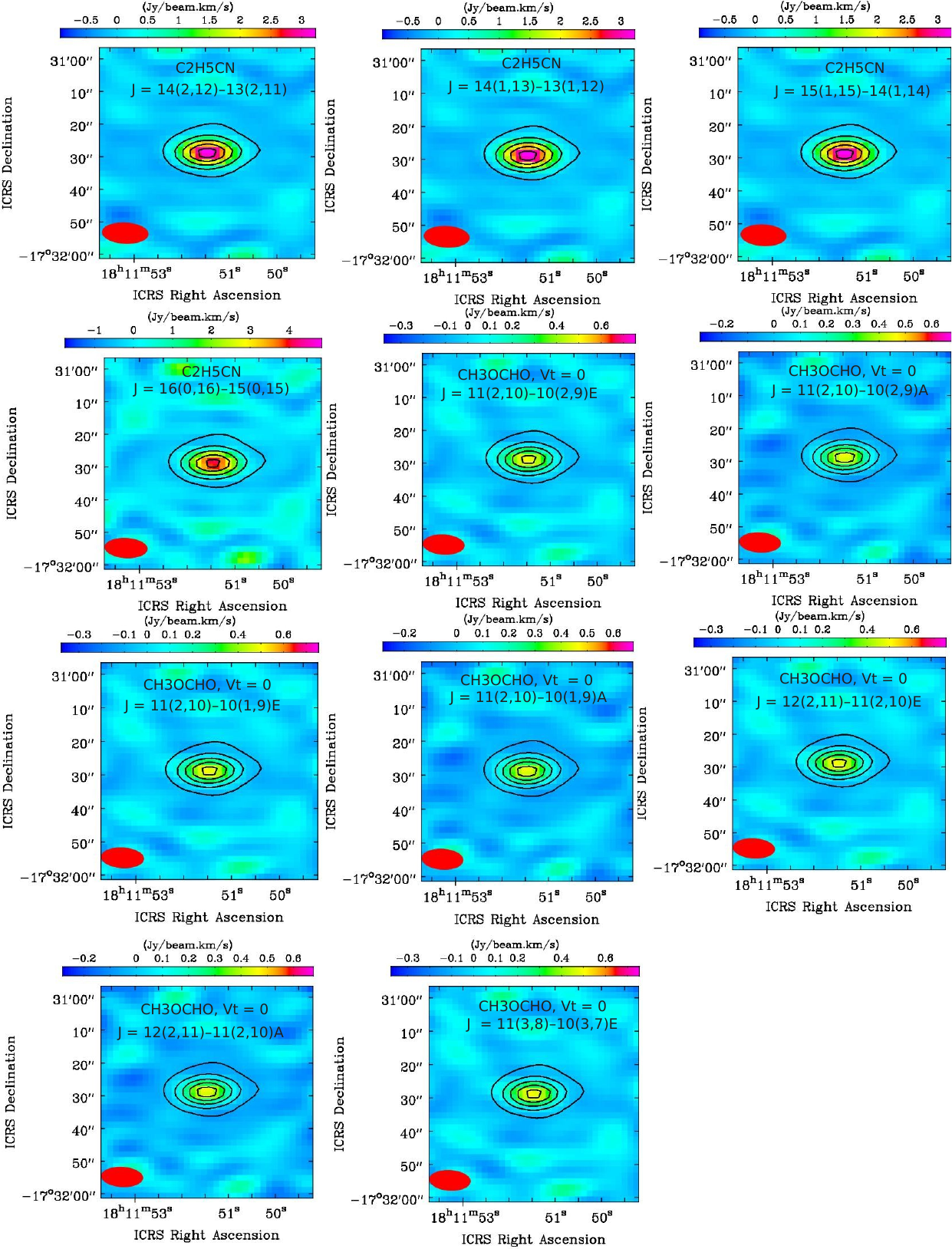}\\
	\caption{Integrated intensity maps of \ce{C2H5CN}, \ce{CH3OCHO} (both V$_{t}$ = 0 and V$_{t}$ = 1), \ce{HC3N} (both $v$7 = 1 and $v$7 = 2), and \ce{NH2CHO} towards the IRAS 18089, which are overlaid with the 2.1 mm continuum emission map of IRAS 18089 (black contour). The contour levels are at 20\%, 40\%, 60\%, 80\%, and 100\% of the peak flux. The red circles represent the synthesised beam of the integrated emission maps.}
	\label{fig:emissionmap}
\end{figure*}

\begin{figure*}
	\text{{\large Figure~5 Continued.}}
	\centering
	\includegraphics[width=1.0\textwidth]{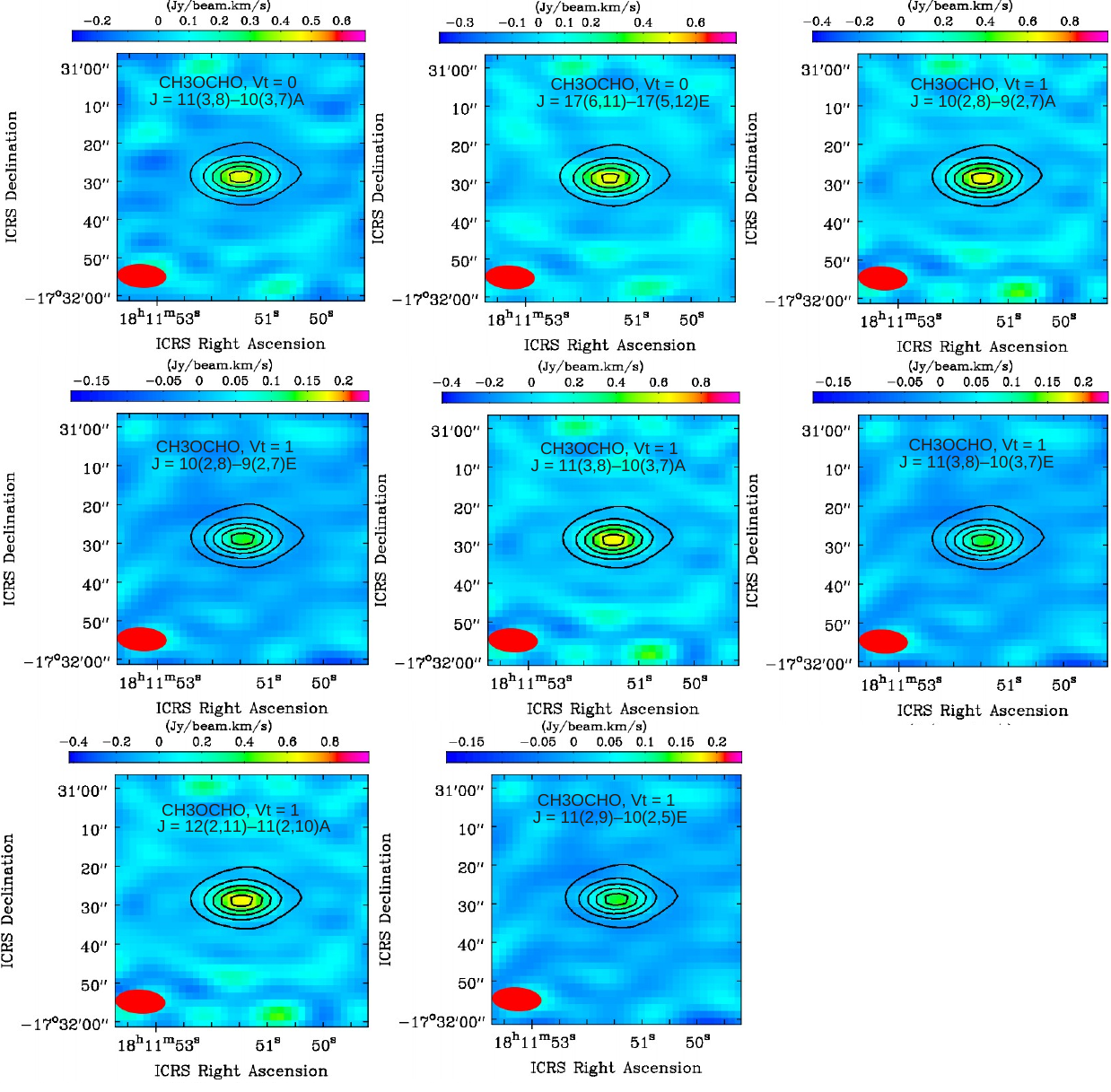}
	\includegraphics[width=1.0\textwidth]{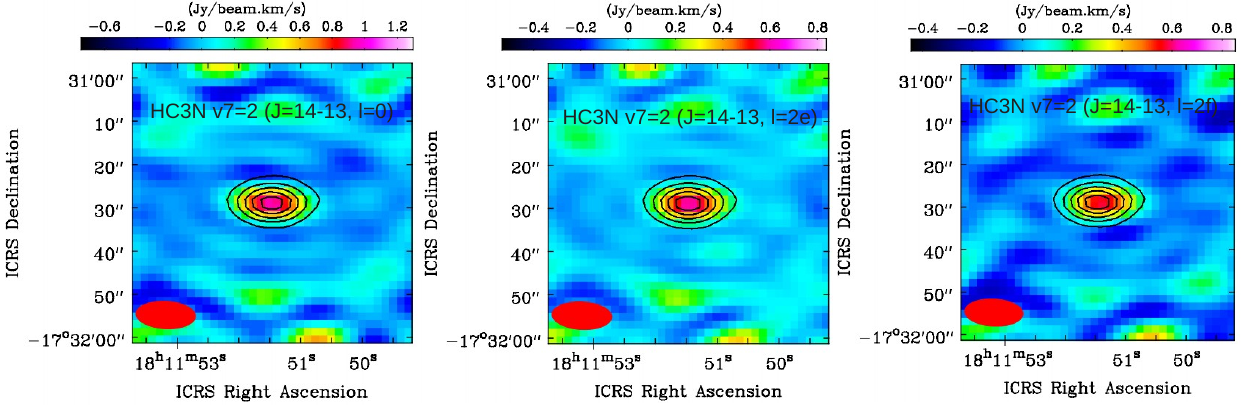}
	\includegraphics[width=1.0\textwidth]{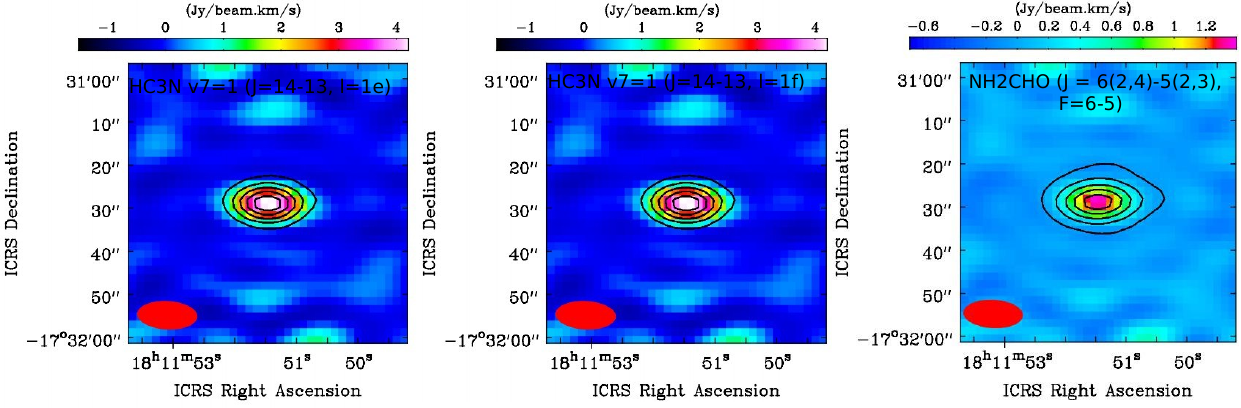}
\end{figure*}

\subsection{Spectral line analysis of IRAS 18089}
\label{coms}
We extracted the molecular spectrum of the high-mass protostar IRAS 18089 for a 20.2$^{\prime\prime}$ diameter circular region centering at ($\alpha,\delta$)$_{\rm J2000}$ = 18:11:51.400, --17:31:28.00. The synthesised beam sizes of the spectral images of IRAS 18089 at frequency ranges of 127.49--128.48 GHz, 129.75--130.75 GHz, 139.09--140.09 GHz, and 140.45--141.45 GHz are 12.82$^{\prime\prime}\times$6.40$^{\prime\prime}$, 12.90$^{\prime\prime}\times$6.32$^{\prime\prime}$, 12.66$^{\prime\prime}\times$6.02$^{\prime\prime}$, and 12.18$^{\prime\prime}\times$6.09$^{\prime\prime}$ respectively. The systemic velocity ($V_{LSR}$) of the IRAS 18089 is 33.8 km s$^{-1}$ \citep{beu05b}. For identification of the emission line of COMs in the spectra, we use local thermodynamic equilibrium (LTE) model spectra with the Cologne Database for Molecular Spectroscopy (CDMS) molecular database \citep{mu05}. For LTE analysis, we use the LTE-RADEX module in CASSIS \citep{vas15}. The LTE assumption is valid in this case as the gas density of the warm inner region of the target is 1.3$\times$10$^{7}$ cm$^{-3}$ \citep{san21}. The identification of the different types of COMs is discussed below.

\subsubsection{Ethyl cyanide (\ce{C2H5CN})}
We detected a total of four rotational emission lines of the complex N-bearing molecule \ce{C2H5CN} towards the IRAS 18089 using the LTE-modelled spectra. The upper-state energies ($E_{u}$) of the detected emission lines of \ce{C2H5CN} vary between 47.23 K and 57.48 K. A Gaussian fit to the observed spectra of \ce{C2H5CN} was performed to estimate the proper FWHM in km s$^{-1}$ and integrated intensity ($\rm{\int T_{mb}dV}$) in K km s$^{-1}$. The resultant LTE and Gaussian fitting spectral lines of \ce{C2H5CN} are shown in Figure~\ref{fig:coms}, and the corresponding spectral parameters are listed in Table~\ref{tab:MOLECULAR DATA1}. We also confirmed that the detected emission lines of \ce{C2H5CN} are non-blended, and all detected emission lines of \ce{C2H5CN} exhibit above 5$\sigma$ significance. The best-fit column density of \ce{C2H5CN} is (1.13$\pm$0.4)$\times$10$^{14}$ cm$^{-2}$ with an excitation temperature of 200$\pm$54 K, which is obtained using the LTE modeling. The fractional abundance of \ce{C2H5CN} towards the IRAS 18089 with respect to \ce{H2} is (1.40$\pm$0.5)$\times$10$^{-10}$.

\subsubsection{Methyl formate (\ce{CH3OCHO})}
We identified a total of sixteen rotational emission lines of \ce{CH3OCHO} towards the IRAS 18089 using the LTE model with more than 5$\sigma$ significance level. The resultant LTE and Gaussian-fitted spectral profiles of \ce{CH3OCHO} are shown in Figure~\ref{fig:coms}, and the spectral line parameters of \ce{CH3OCHO} are presented in Table~\ref{tab:MOLECULAR DATA1}. We observed the V$_{t}$ = 0 and V$_{t}$ = 1 states in the transitions of the \ce{CH3OCHO}. The V$_{t}$ = 0 and V$_{t}$ = 1 states indicate the ground and first torsionally excited states of \ce{CH3OCHO}. We find that the upper state energies ($E_{u}$) of the detected emission lines of \ce{CH3OCHO} (V$_{t}$ = 0) vary between 40.71 K and 114.60 K. Similarly, the upper state energies ($E_{u}$) of the \ce{CH3OCHO} (V$_{t}$ = 1) vary between 223.32 K and 235.0 K. We also observe the torsional substates A and E in the transitions of the \ce{CH3OCHO} due to the internal rotation of the methyl (\ce{CH3}) group. During the spectral fitting using the LTE model, we observed that the J = 17(6,11)--17(5,12)A transition line of \ce{CH3OCHO} (V$_{t}$ = 0) is not fitted well because that emission line is blended with \ce{CH3OCH3}. In the Table~\ref{tab:MOLECULAR DATA1}, the transitions of \ce{CH3OCHO} are presented corresponding to V$_{t}$ = 0 and V$_{t}$ = 1 states. Using the LTE model, the best-fit column density of the \ce{CH3OCHO} V$_{t}$ = 0 is (1.5$\pm$0.3)$\times$10$^{15}$ cm$^{-2}$ and V$_{t}$ = 1 is (7.1$\pm$0.5)$\times$10$^{14}$ cm$^{-2}$ with excitation temperatures of 200$\pm$48 K and 196$\pm$74 K. The column density of \ce{CH3OCHO} varies between the V$_{t}$ = 0 and V$_{t}$ = 1 torsional excited states because the first torsional excited state (V$_{t}$ = 1) of \ce{CH3OCHO} is located at about 132 cm$^{-1}$ above the ground state (V$_{t}$ = 0), where 1 cm$^{-1}$ = 1.438 K \citep{tu12}. The estimated abundances of \ce{CH3OCHO} with respect to \ce{H2} towards the IRAS 18089 are (1.90$\pm$0.9)$\times$10$^{-9}$ (for V$_{t}$ = 0) and (8.90$\pm$0.8)$\times$10$^{-10}$ (for V$_{t}$ = 1).

\begin{table*}
	%	\begin{minipage}[t]{\columnwidth}
	\centering
	%\scriptsize 
	\caption{Summary of the line properties of the detected molecules towards the IRAS 18089.}
	\begin{adjustbox}{width=1.0\textwidth}
		\begin{tabular}{|c|c|c|c|c|c|c|c|c|c|c|c|c|c|c|c|c|} 
			\hline 
			Molecule&Observed frequency$^{\dagger}$ &Transition$^{\dagger}$ & $E_{u}$$^{\dagger}$ & $A_{ij}$$^{\dagger}$&FWHM$^{*}$& $\rm{\int T_{mb}dV}$$^{*}$&Column density& Excitation temperature& Remark$^{\bullet}$\\
			
			&(GHz) & &(K)&(s$^{-1}$)&(km s$^{-1}$)&(K km s$^{-1}$)&(cm$^{-2}$)& (K) & \\
			\hline
			C$_{2}$H$_{5}$CN&127.520&14(2,12)--13(2,11)&50.02&1.69$\times$10$^{-4}$ &5.52$\pm$0.24 &0.587& & &Non blended\\
			\cline{2-7}
			\cline{10-10}
			
			&127.618&14(1,13)--13(1,12)&47.23&1.72$\times$10$^{-4}$&5.16$\pm$0.28 & 0.563&(1.13$\pm$0.4)$\times$10$^{14}$ cm$^{-2}$ &200$\pm$54 K  &Non blended\\
			\cline{2-7}
			\cline{10-10}
			&129.795&15(1,15)--14(1,14)&51.09&1.82$\times$10$^{-4}$&4.82$\pm$0.64 &0.255& & &Non blended\\
			\cline{2-7}
			\cline{10-10}
			&139.335&16(0,16)--15(0,15)&57.48&2.26$\times$10$^{-4}$&5.16$\pm$0.22 &0.621& & &Non blended\\
			\hline
			CH$_{3}$OCHO, $v_{t}$ = 0&130.010&11(2,10)--10(2,9)E&40.71&3.12$\times$10$^{-5}$&3.36$\pm$0.35 &0.284& & &Non blended\\
			\cline{2-7}	
			\cline{10-10}
			&130.016&11(2,10)--10(2,9)A&40.70&3.12$\times$10$^{-5}$&3.56$\pm$0.42 &0.293& & &Non blended\\	
			\cline{2-7}
			\cline{10-10}
			&139.565&11(2,10)--10(1,9)E&40.71&3.82$\times$10$^{-6}$&3.10$\pm$0.36 &0.063 & & &Non blended\\
			\cline{2-7}	
			\cline{10-10}
			&139.569&11(2,10)--10(1,9)A&40.70&3.80$\times$10$^{-6}$&3.23$\pm$0.28 &0.081 & & &Non blended\\
			\cline{2-7}
			\cline{10-10}
			&141.037&12(2,11)--11(2,10)E&47.48&4.01$\times$10$^{-5}$&3.72$\pm$0.29 &0.784 &(1.5$\pm$0.3)$\times$10$^{15}$ cm$^{-2}$&  200$\pm$48 K  &Non blended\\
			\cline{2-7}
			\cline{10-10}
			&141.044&12(2,11)--11(2,10)A&47.47&4.01$\times$10$^{-5}$&4.22$\pm$0.89 &0.875 & & &Non blended\\
			\cline{2-7}
			\cline{10-10}
			&141.244&11(3,8)--10(3,7)E&47.77&3.87$\times$10$^{-5}$&3.90$\pm$0.31 &0.995 & & &Non blended\\
			\cline{2-7}
			\cline{10-10}
			&141.260&11(3,8)--10(3,7)A&47.75&3.87$\times$10$^{-5}$&4.01$\pm$0.35 &0.846 & & &Non blended\\
			\cline{2-7}
			\cline{10-10}
			&141.350&17(6,11)--17(5,12)A&114.59&4.09$\times$10$^{-6}$&5.86$\pm$1.09 &0.199 & & &Blended with CH$_{3}$OCH$_{3}$\\
			\cline{2-7}
			\cline{10-10}
			&141.354&17(6,11)--17(5,12)E&114.60&3.98$\times$10$^{-6}$&3.37$\pm$0.50 &0.054 & & &Non blended\\
			\hline
			CH$_{3}$OCHO, $v_{t}$ = 1&127.836&10(2,8)--9(2,7)A&223.96&2.95$\times$10$^{-5}$&4.59$\pm$0.78 &0.102 & & &Non blended\\
			\cline{2-7}
			\cline{10-10}
			&128.273&10(2,8)--9(2,7)E&223.32&2.99$\times$10$^{-5}$&3.96$\pm$0.69 &0.076 & & &Non blended\\
			\cline{2-7}
			\cline{10-10}
			&139.381&11(3,8)--10(3,7)A&233.19&3.70$\times$10$^{-5}$&4.63$\pm$0.86 &0.133 &(7.1$\pm$0.5)$\times$10$^{14}$ cm$^{-2}$& 196$\pm$74 K &Non blended\\
			\cline{2-7}
			\cline{10-10}
			&139.635&11(3,8)--10(3,7)E&232.67&3.72$\times$10$^{-5}$&4.36$\pm$0.44 &0.124 & & &Non blended\\
			\cline{2-7}
			\cline{10-10}
			&140.020&12(2,11)--11(2,10)A&235.00&3.92$\times$10$^{-5}$&4.05$\pm$0.23 &0.144 & & &Non blended\\
			\cline{2-7}
			\cline{10-10}
			&140.622&11(2,9)--10(2,5)E&230.07&3.98$\times$10$^{-5}$&4.81$\pm$0.73 &0.253 & & &Non blended\\
			\hline
			\ce{HC3N}, $v$7=1&127.680&J = 14--13, $l$ = 1$e$&366.80&1.61$\times$10$^{-4}$&7.22$\pm$0.30&1.258 &(6.02$\pm$0.5)$\times$10$^{13}$ cm$^{-2}$& 160$\pm$25 K  &Non blended\\
			\cline{2-7}
			\cline{10-10}
			&127.863&J = 14--13, $l$ = 1$f$&366.86&1.62$\times$10$^{-4}$&6.98$\pm$0.22&1.072& & &Non blended\\		
			
			\hline
			\ce{HC3N}, $v$7=2&128.169&J = 14--13, $l$ = 0&687.82&1.63$\times$10$^{-4}$&3.61$\pm$0.53 &0.112 &(2.50$\pm$0.3)$\times$10$^{13}$ cm$^{-2}$& 180$\pm$20 K &Non blended\\
			\cline{2-7}
			\cline{10-10}
			&128.175&J = 14--13, $l$ = 2$e$&691.09&1.60$\times$10$^{-4}$&3.89$\pm$0.82 &0.233 & & &Non blended\\
			\cline{2-7}
			\cline{10-10}
			&128.182&J = 14--13, $l$ = 2$f$&691.10&1.60$\times$10$^{-4}$&3.96$\pm$0.68 &0.349 & & &Non blended\\
			
			\hline
			\ce{NH2CHO}&128.102&J = 6(2,4)--5(2,3), F = 6--5&33.38&1.28$\times$10$^{-4}$&5.20$\pm$0.86&0.429 &(5.01$\pm$0.2)$\times$10$^{13}$ cm$^{-2}$ & 180$\pm$62 K &Non blended\\
			\hline
		\end{tabular}	
	\end{adjustbox}
	\\
	${\dagger}$--Frequency, transitions, $E_{u}$, A$_{ij}$ are obtained from the LTE modelling.\\
	${*}$ -- FWHM and $\rm{\int T_{mb}dV}$ are an estimate from the Gaussian fitting over the molecular lines.\\
	${\bullet}$--Blended and non-blended effects in the observed spectral lines are verified by the LTE modelling and online molecular line list database {Splatalogue}.\\
	\label{tab:MOLECULAR DATA1}
	%	\end{minipage}[t]{\columnwidth}
\end{table*}

\subsubsection{Cyanoacetylene (\ce{HC3N})}
We detected two vibrationally excited states of \ce{HC3N} ($v$7 = 1 and $v$7 = 2) towards the IRAS 18089. Using the LTE-modelled spectra, we identified two rotational emission lines of $v$7 = 1 state and three rotational emission lines of $v$7 = 2 states of \ce{HC3N}. The detected vibrationally excited \ce{HC3N} emission lines exhibit the higher upper-state energies of $E_{up}$/k $\sim$ 366.80--366.86 K (for $v$7 = 1) and $\sim$ 687.82--691.10 K (for $v$7 = 2). The higher vibrational states of \ce{HC3N} ($v$7 = 1 and $v$7 = 2) with higher upper state energies have the ability to trace the inner hot gas near the central star. The resultant detected spectral lines of \ce{HC3N} ($v$7 = 1 and $v$7 = 2) are shown in Figure~\ref{fig:coms}, and the spectral line parameters of \ce{HC3N} ($v$7 = 1 and $v$7 = 2) are presented in Table~\ref{tab:MOLECULAR DATA1}. Using the LTE model, the best-fit column density of the \ce{HC3N} with $v$7 = 1 and $v$7 = 2 excited states are (6.02$\pm$0.5)$\times$10$^{13}$ cm$^{-2}$ and (2.50$\pm$0.3)$\times$10$^{13}$ cm$^{-2}$ respectively, with excitation temperatures of 160$\pm$25 K and 180$\pm$20 K. The estimated fractional abundances of \ce{HC3N} towards the IRAS 18089 are (7.5$\pm$0.7)$\times$10$^{-11}$ (for $v$7 = 1) and (3.1$\pm$0.4)$\times$10$^{-11}$ (for $v$7 = 2).

\subsubsection{Formamide(\ce{NH2CHO})}
We have detected a single transition line of \ce{NH2CHO} towards IRAS 18089 using the LTE-modelled spectra. We detected the J = 6(2,4)--5(2,3), F = 6--5 transition line of \ce{NH2CHO}. We do not observe any other transition lines of \ce{NH2CHO} in the entire data. The upper-state energy ($E_{u}$) of the detected emission line of \ce{NH2CHO} is 33.38 K. To estimate the proper FWHM in km s$^{-1}$ and integrated intensity ($\rm{\int T_{mb}dV}$) in K km s$^{-1}$ of the emission line of \ce{NH2CHO}, we fitted the Gaussian model over the observed spectra of \ce{NH2CHO}. The resultant LTE and Gaussian fitting spectral lines of \ce{NH2CHO} are shown in Figure~\ref{fig:coms}, and the corresponding spectral parameters are listed in Table~\ref{tab:MOLECULAR DATA1}. We observed that the detected emission lines of \ce{NH2CHO} are non-blended and exhibit above 5$\sigma$ significance. Using the LTE spectral modelling, the best-fit column density of \ce{NH2CHO}, based on the single transition line, is (5.01$\pm$0.2)$\times$10$^{13}$ cm$^{-2}$ with an excitation temperature of 180$\pm$62 K. The fractional abundance of \ce{NH2CHO} with respect to \ce{H2} towards the IRAS 18089 is (6.25$\pm$0.8)$\times$10$^{-11}$.

\begin{table*}
	%	\begin{minipage}[t]{\columnwidth}
	\centering
	\scriptsize 
	\caption{Estimated emitting regions of detected COMs towards the IRAS 18089.}
	\begin{adjustbox}{width=0.8\textwidth}
		\begin{tabular}{ccccccccccccccccc}
			\hline 
			Molecule&Observed frequency&Transition& Emitting region & Physical scale$^{\dagger}$\\
			&(GHz) & &($^{\prime\prime}$)& (pc) \\
			\hline
			C$_{2}$H$_{5}$CN&127.520&14(2,12)--13(2,11)&12.09$\pm$0.61&29.43\\
			&127.618&14(1,13)--13(1,12)&12.10$\pm$0.65&29.45\\
			&129.795&15(1,15)--14(1,14)&12.12$\pm$0.59&29.50\\
			&139.335&16(0,16)--15(0,15)&12.11$\pm$0.60&29.47\\
			\hline
			CH$_{3}$OCHO, $v_{t}$ = 0&130.010&11(2,10)--10(2,9)E&12.08$\pm$0.58&29.40\\	
			&130.016&11(2,10)--10(2,9)A&12.07$\pm$0.57&29.38\\	
			&139.565&11(2,10)--10(1,9)E&12.09$\pm$0.59&29.43\\	
			&139.569&11(2,10)--10(1,9)A&12.11$\pm$0.61&29.47\\
			&141.037&12(2,11)--11(2,10)E&12.14$\pm$0.58&29.55\\
			&141.044&12(2,11)--11(2,10)A&12.09$\pm$0.55&29.43\\
			&141.244&11(3,8)--10(3,7)E&12.15$\pm$0.59&29.57\\
			&141.260&11(3,8)--10(3,7)A&12.11$\pm$0.58&29.47\\
			&141.354&17(6,11)--17(5,12)E&12.08$\pm$0.61&29.40\\
			\hline
			CH$_{3}$OCHO, $v_{t}$ = 1&127.836&10(2,8)--9(2,7)A&12.11$\pm$0.58&29.48\\
			&128.276&10(2,8)--9(2,7)E&12.10$\pm$0.61&29.45\\		
			&139.381&11(3,8)--10(3,7)A&12.14$\pm$0.59&29.55\\
			&139.635&11(3,8)--10(3,7)E&12.07$\pm$0.57&29.38\\
			&140.020&12(2,11)--11(2,10)A&12.11$\pm$0.60&29.47\\
			&140.622&11(2,9)--10(2,5)E&12.09$\pm$0.56&29.43\\
			
			\hline
			\ce{HC3N}, $v$7=1&127.680&J = 14--13, $l$ = 1$e$&12.11$\pm$0.62&29.47\\
			&127.863&J = 14--13, $l$ = 1$f$&12.12$\pm$0.63&29.50\\			
			
			\hline
			
			\ce{HC3N}, $v$7=2&128.169&J = 14--13, $l$ = 0&12.10$\pm$0.58&29.45\\
			&128.175&J = 14--13, $l$ = 2$e$&12.11$\pm$0.60&29.47\\
			&128.182&J = 14--13, $l$ = 2$f$&12.10$\pm$0.59&29.45\\
			\hline  
			\ce{NH2CHO}&128.102&J = 6(2,4)--5(2,3), F = 6--5&12.13$\pm$0.62&29.52\\
			\hline
		\end{tabular}	
	\end{adjustbox}
	\label{tab:emitting region}\\
	%	\end{minipage}[t]{\columnwidth}
	$\dagger$--During the estimation of the physical scale of the emitting regions of different molecules, we used the cosmological parameters using Plank with Hubble constant $H_{0}$ = 0.6731, $\Omega_{M}$ = 0.315, and $\Omega_{L}$ = 0.685 \citep{ag20}.
\end{table*}

\subsection{Spatial distribution of detected COMs}
We create the integrated intensity maps of the non-blended transitions of \ce{C2H5CN}, \ce{CH3OCHO} (V$_{t}$ = 0 and V$_{t}$ = 1), \ce{HC3N} ($v$7 = 1 and $v$7 = 2), and \ce{NH2CHO} using the CASA task {\tt IMMOMENTS} towards the IRAS 18089. The integrated intensity maps of detected COMs towards the IRAS 18089 are created by integrating the spectral data cubes over carefully determined channel ranges where COMs are detected (see Figure~\ref{fig:emissionmap}). We notice that the integrated intensity maps of different COMs have a peak at the position of the continuum. From the integrated emission maps it is clear that the emission lines of different COMs may have arised from the extended envelope of the IRAS 18089. We also estimated the emitting regions of COMs towards the IRAS 18089 by fitting the 2D Gaussian over the integrated intensity maps using the CASA {\tt IMFIT} task. The deconvolved beam size of the emitting region of COMs is estimated by using the following equation,\\

$\theta_{S}=\sqrt{\theta^2_{50}-\theta^2_{beam}}$~~~~~~~~~~~~~~~~~~~~~~~~~~(5)\\\\
where $\theta_{50} = 2\sqrt{A/\pi}$ denotes the diameter of the circle with area, $A$ is the area that encloses 50\% of the peak intensity, and $\theta_{beam}$ is the half-power width of the synthesised beam of the integrated emission maps \citep{riv17}. The value of $\theta_{50}$ is estimated using the task {\tt IMFIT}. The estimated emitting regions of COMs are shown in Table~\ref{tab:emitting region}. The emitting regions of \ce{C2H5CN} vary between 12.09$^{\prime\prime}$ (29.43 pc)--12.12$^{\prime\prime}$ (29.50 pc). The emitting regions of \ce{CH3OCHO} with V$_{t}$ = 0 and V$_{t}$ = 1 states vary between 12.07$^{\prime\prime}$ (29.38 pc)--12.15$^{\prime\prime}$ (29.57 pc) and 12.07$^{\prime\prime}$ (29.38 pc)--12.14$^{\prime\prime}$ (29.55 pc). The emitting regions of \ce{HC3N} with vibrationally excited states $v$7 = 1 and $v$7 = 2 vary between 12.11$^{\prime\prime}$ (29.47 pc)--12.12$^{\prime\prime}$ (29.50 pc) and 12.10$^{\prime\prime}$ (29.45 pc)--12.11$^{\prime\prime}$ (29.47 pc). Similarly, the emitting region of \ce{NH2CHO} is 12.13$^{\prime\prime}$ (29.52 pc). A smaller emitting region of COMs in comparison to the synthesised beam size implies that the emitting area of COMs is spatially unresolved in the current data set. Further high-resolution observations of these lines are required, possibly using an ALMA 12-m array, to examine the distribution and chemical morphology of COMs in the IRAS 18089 region.

\section{Discussion}
\label{dis}
\subsection{Possible formation mechanism of COMs}
\label{formation}
\subsubsection{Ethyl cyanide (\ce{C2H5CN}):}
The N-bearing molecule, \ce{C2H5CN}, is formed on the grain surface of the high-mass protostars and hot molecular cores \citep[see ][for details]{meh04, gar13, gar17, gar22}. \citet{gar13} showed that the subsequential hydrogenation of \ce{HC3N} in the free-fall collapse phase can form the vinyl cyanide (\ce{CH2CHCN}) (\ce{HC3N} + 2H $\rightarrow$ \ce{CH2CHCN}). The subsequential hydrogenation of the \ce{CH2CHCN} form \ce{C2H5CN} in the grain surface of the hot molecular cores and high-mass protostars \citep[\ce{CH2CHCN} + 2H $\rightarrow$ \ce{C2H5CN};][]{gar13}, through barrierless and exothermic radical-radical reactions \citep{sin21}. This particular chemical reaction is demonstrated to be the most efficient way for the formation of \ce{C2H5CN} towards the Sgr B2, G10.47+0.03, G31.41+0.31, and other molecular cores \citep{bel09, man23, min23}. This reaction may be most efficient towards IRAS 18089 because \ce{HC3N} acts as a possible precursor of \ce{C2H5CN}, and we detected the emission lines of \ce{HC3N} from this source, which we already discussed in this paper.

\subsubsection{Methyl formate (\ce{CH3OCHO}):}
In this article, we present the first detection of \ce{CH3OCHO} with V$_{t}$ = 0 and V$_{t}$ = 1 states towards the IRAS 18089 using the ACA. In the hot molecular cores and high-mass protostars, \ce{CH3OCHO} molecule can efficiently be formed by the reaction of \ce{CH3O} and HCO radicals on the surface of dust grains (\ce{CH3O} + HCO $\rightarrow$ \ce{CH3OCHO}) \citep[see][and references therein] {gar08}. Earlier, \citet{man22} showed that the reactions between radical \ce{CH3O} and radical HCO produce \ce{CH3OCHO} towards the hot molecular core IRAS 18566+0408. \citet{gar13} showed the chemical reaction between \ce{CH3O} and \ce{HCO} is mobile between 30--40 K, and this chemical reaction is the most efficient pathway to the formation of \ce{CH3OCHO} towards the hot molecular cores and high-mass protostars. According to the chemical modelling of \citet{gar13}, the gas phase \ce{CH3OCHO} mainly comes from the ice phase of massive protostars. Earlier, \citet{gor21} reported that the UV photodissociation of \ce{CH3OH} leads to the formation of \ce{CH2O}, \ce{CH3O}, and \ce{CH3} at around 40 K temperature, and these molecules create the COMs, like \ce{CH3OCHO}, \ce{CH3OCH3}. At temperature $T\sim$ 40 K in the gas phase, the reaction of protonated \ce{CH3OH} and \ce{H2CO} creates \ce{H5C2O2}$^{+}$ \citep{gor21}. The \ce{CH3OCHO} is created in the hot molecular cores and massive protostars via the electron recombination of \ce{H5C2O2}$^{+}$ (\ce{H5C2O2}$^{+}$+e$^{-}\longrightarrow$ \ce{CH3OCHO}+H) \citep{bon19, gor21}. Earlier, \citet{bal15} proposed an efficient gas phase reaction of \ce{CH3OCHO} in a cold environment where \ce{CH3OCH3} behaves as a possible precursor of \ce{CH3OCHO}. Therefore, it is evident that \ce{CH3OCHO} can be produced through both grain surface and gas phase reactions in the hot molecular cores and high-mass protostars. Note that \cite{beu05b} detected \ce{CH3OCH3} with transition J = 19(1,18)--18(2,17)AA towards IRAS 18089 using SMA. However, such detection of a single transition line of one complex molecule could not be considered as conclusive evidence of the presence of the molecule in high-mass protostars.

\subsubsection{Cyanoacetylene (\ce{HC3N}):}
In the ISM, acetylene (\ce{C2H2}) is one of the known compounds that is formed in the grain surfaces \citep{ch09}. After the evaporation of \ce{C2H2} in the grains, there are two possible scenarios. \\
1. When the UV field is strong, like in photodissociation regions (PDR), then \ce{C2H2} photo-dissociates into the ethynyl radical (\ce{C2H}) (\ce{C2H2} + $h\nu$$\rightarrow$ \ce{C2H} + H) \citep{ch93, me05}.\\
2. In the weak UV field regime (without PDR) in the presence of cyano radical (CN), \ce{C2H2} reacts with the CN to create \ce{HC3N} (\ce{C2H2} + CN $\rightarrow$ \ce{HC3N} + H) \citep{fu97, me05, ch09}. Recently, \cite{tan22} claimed that the \ce{HC3N} is created in the high-mass protostars and UC H{\sc ii} regions when \ce{C2H2} immediately reacts with the CN. The \ce{HC3N} molecule is associated with highly dense, warm, and shielded gas in the star-forming regions or high-mass protostars \citep{tan22}.

In the star-forming regions, with the presence of UV radiation, the \ce{HC3N} molecule is easily destroyed by reacting with C+ ions \citep{rod98, me05}. After the destruction, \ce{HC3N} converts into \ce{C2H} or \ce{C3N} if \ce{HC3N} is photo-dissociated \citep{ch93}. Similarly, \ce{HC3N} converts into \ce{C3H+} or \ce{C4N+} if \ce{HC3N} reacts with C+ \citep{bo85}. The possible destruction reactions are:\\\\
\ce{HC3N} + $h\nu$ $\rightarrow$ \ce{C2H} + CN~~~~~~~~~~~~~~(1)\\\\
\ce{HC3N} + $h\nu$ $\rightarrow$ \ce{C3N} + H~~~~~~~~~~~~~~~~~(2)\\\\
\ce{HC3N} + C$^{+}$ $\rightarrow$ \ce{C3H}$^{+}$ + CN~~~~~~~~~~~(3)\\\\
\ce{HC3N} + C$^{+}$ $\rightarrow$ \ce{C4N}$^{+}$ + H~~~~~~~~~~~~~~(4)\\\\
The rates of the above reaction are reported in \citet{ch93}, and \cite{tan22} showed that these destruction pathways are the most efficient towards the high-mass protostars and UC H{\sc ii} regions.

\subsubsection{Formamide (\ce{NH2CHO})}
In the ISM, the formation pathways of \ce{NH2CHO} are still a topic of debate in both gas phase and solid state ice mantle formation routes \citep{gar08}. Previously, \citet{gar13} showed that \ce{NH2CHO} is created in the gas phase barrierless reaction between \ce{NH2} and \ce{H2CO} (\ce{NH2} + \ce{H2CO} $\rightarrow$ \ce{NH2CHO} + H). Earlier, \citet{gar13} and \cite{suz18} showed that reaction is the most efficient for the formation of \ce{NH2CHO} towards hot molecular cores. After that, \cite{gor20} claimed that reaction is responsible for the production of \ce{NH2CHO} towards G10.47+0.03.

\subsection{Comparision between the observed and modelled abundance of COMs}
Here we compare the estimated fractional abundances of \ce{C2H5CN}, \ce{CH3OCHO}, \ce{HC3N}, and \ce{NH2CHO} towards the IRAS 18089 with the existing three-phase warm-up chemical model abundances of \cite{suz18}. For chemical modelling, \cite{suz18} used the gas-grain chemical kinetics code NAUTILUS in the environment of hot molecular cores and massive protostars. During the chemical modelling of complex molecules, \cite{suz18} assumed an isothermal collapse phase after a static warm-up phase. In the first phase, the gas density increases from 3$\times$10$^{3}$ cm$^{-3}$ to 2$\times$10$^{7}$ cm$^{-3}$ and dust temperature decreases from 16 K to 8 K under the free-fall collapse \citep{suz18}. In the second phase, known as the warm-up phase, the dust temperature increases from 8 to 400 K, and the gas density remains fixed at 2$\times$10$^{7}$ cm$^{-3}$. For the chemical modelling, \cite{suz18} used the following reactions in the grain surfaces to produce the \ce{C2H5CN}, \ce{CH3OCHO}, and \ce{HC3N}.\\\\
\ce{CH2CHCN} + 2H $\rightarrow$ \ce{C2H5CN}~~~~~~~~~~~~~~~~~~~~(5)\\\\
\ce{CH3O} + HCO $\rightarrow$ \ce{CH3OCHO}~~~~~~~~~~~~~~~~~~~~~(6)\\\\
\ce{C2H2} + CN $\rightarrow$ \ce{HC3N} + H~~~~~~~~~~~~~~~~~~~~~~~~~~~(7)\\\\
\ce{NH2} + \ce{H2CO} $\rightarrow$ \ce{NH2CHO} + H~~~~~~~~~~~~~~~~~~(8)\\\\
We explain reactions 5, 6, 7, and 8 in Section~\ref{formation}. The three-phase warm-up chemical modelling of \cite{suz18} is appropriate for explaining the chemical abundance and evolution of COMs towards the IRAS 18089 because the gas density of IRAS 18089 is 1.3$\times$10$^{7}$ cm$^{-3}$ \citep{san21} and the gas temperature is above 100 K \citep{beu041}. After the chemical modelling, \cite{suz18} determined the modelled abundance of \ce{C2H5CN}, \ce{CH3OCHO}, \ce{HC3N}, and \ce{NH2CHO} to be 6.3$\times$10$^{-10}$, 4.9$\times$10$^{-9}$, 1.1$\times$10$^{-11}$, and 4.0$\times$10$^{-11}$ respectively using the fast warm-up model with the help of reactions 5, 6, 7, and 8 on the grain surfaces and gas phases of hot molecular cores and massive prostars. Our estimated abundances of \ce{C2H5CN}, \ce{CH3OCHO} (V$_{t}$ = 0), \ce{HC3N}, and \ce{NH2CHO} is shown in Section~\ref{coms}. We find that our estimated abundances of \ce{C2H5CN}, \ce{CH3OCHO}, \ce{HC3N}, and \ce{NH2CHO} are nearly similar to the chemical modelling abundances of \cite{suz18}. This result indicates that reactions 5, 6, 7, and 8 may be responsible for the production of \ce{C2H5CN}, \ce{CH3OCHO}, \ce{HC3N}, and \ce{NH2CHO} towards the IRAS 18089. 

\section{Conclusion}
\label{con}	
In this article, we present the identification of the rotational emission lines of \ce{C2H5CN}, \ce{CH3OCHO} (V$_{t}$ = 0 and V$_{t}$ = 1), \ce{HC3N} ($v$7 = 1 and $v$7 = 2), and \ce{NH2CHO} towards the IRAS 18089. The estimated abundance of \ce{C2H5CN} towards the IRAS 18089 is (1.40$\pm$0.5)$\times$10$^{-10}$. The estimated abundance of \ce{CH3OCHO} with respect to \ce{H2} is (1.90$\pm$0.9)$\times$10$^{-9}$ and (8.90$\pm$0.8)$\times$10$^{-10}$ for V$_{t}$ = 0 and V$_{t}$ = 1, respectively. The abundance of \ce{HC3N} towards the IRAS 18089 is (7.5$\pm$0.7)$\times$10$^{-11}$ (for $v$7 = 1) and (3.1$\pm$0.4)$\times$10$^{-11}$ (for $v$7 = 2). Similarly, the abundance of \ce{NH2CHO} towards IRAS 18089 is (6.25$\pm$0.8)$\times$10$^{-11}$. We created the integrated emission maps of detected COMs, and we observed that the detected emission lines of different COMs arise from the extended envelope of the protostar. We also compare the estimated abundances of \ce{C2H5CN}, \ce{CH3OCHO}, \ce{HC3N}, and \ce{NH2CHO} with the existing three-phase warm-up modelling abundances of those molecules. After the comparisons, we notice that the observed and modelled abundances are nearly similar. So, the N-bearing molecules \ce{C2H5CN} and \ce{HC3N} are most probably created via the subsequential hydrogenation of the \ce{CH2CHCN} and the reactions between \ce{C2H2} and CN on the grain surface of IRAS 18089. The O-bearing molecule \ce{CH3OCHO} is probably created via the reactions between radical \ce{CH3O} and radical HCO on the grain surface of IRAS 18089. Similarly, the O- and N-bearing molecule \ce{NH2CHO} is most probably formed between the reactions of \ce{NH2} and \ce{H2CO} in the gas phase of IRAS 18089. The prebiotic chemistry of \ce{C2H5CN}, \ce{CH3OCHO}, \ce{HC3N}, and \ce{NH2CHO} towards the IRAS 18089 suggests that both gas phase and grain surface chemistry are efficient for the production of other COMs in that protostar, including those molecules that are chemically connected with those detected molecules. The detections of both N- and O-bearing molecules towards IRAS 18089 indicate that IRAS 18089 is a reservoir of several other COMs. A spectral line survey is needed using the ALMA 12-m arrays to understand the prebiotic chemistry in this high-mass protostar, which will be carried out in our follow-up study.

\section*{acknowledgements} 
We thank the anonymous referee for the helpful comments that improved the manuscript. A.M. acknowledges the Swami Vivekananda Merit-cum-Means Scholarship (SVMCM) for financial support for this research. This paper makes use of the following ALMA data: ADS /JAO.ALMA\#2016.2.00005.S. ALMA is a partnership of ESO (representing its member states), NSF (USA), and NINS (Japan), together with NRC (Canada), MOST and ASIAA (Taiwan), and KASI (Republic of Korea), in cooperation with the Republic of Chile. The Joint ALMA Observatory is operated by ESO, AUI/NRAO, and NAOJ.

\section*{Conflicts of interest}
The authors declare no conflict of interest.

\bibliographystyle{aasjournal}
%\bibliography{./literature.bib,added.bib} % if your bibtex file is called example.bib

\end{document}